# Non-ideal torque control of wind turbine systems: Impacts on annual energy production

Christoph M. Hackl$^{\star}$ and Korbinian Schechner$^{\dagger}$
Munich School of Engineering, Research group “Control of renewable energy systems (CRES)”
Technische Universität München, Boltzmannstr. 15, 85748 Garching, Germany
e-mail$^{\star}$: christoph.hackl@tum.de

***Abstract*—We discuss non-ideal torque control in wind turbine systems (WTS). Most high-level controllers generate a reference torque which is then send to the underlying electrical drive system (generator+inverter) of the WTS to steer the turbine/generator to its optimal operation point (depending on the wind speed). The energy production heavily depends on the mechanical power (i.e. the product of rotational speed and generator torque). However, since torque sensors in the MW range are not available or extremely expensive, the torque controllers are implemented as feedforward controllers and, therefore, are inherently sensitive to parameter variations/uncertainties. Based on real wind data and a dynamical WTS model, we discuss causes and impacts of non-ideal (feedforward) torque control on the energy production and the gross earnings.**

## Notation

$\mathbb{N}$, $\mathbb{R}$: natural, real numbers. $\boldsymbol{x} := (x_1, \ldots, x_n)^\top \in \mathbb{R}^n$: column vector, $n \in \mathbb{N}$ where ‘$^\top$’ and ‘:=’ mean ‘transposed’ and ‘is defined as’, respectively.

## I. Motivation and introduction

Onshore wind turbine systems (WTS) are already today a competitive alternative to classical power generation with respect to the levelized costs of electricity (LCOE) [1]. The major part of the LCOE of wind power is fixed by the initial investment costs. So, a crucial factor to reduce the LCOE is to maximize the energy production (per year or lifetime). One reason for a non-optimal energy production is *non-ideal torque control*. Since WTS are *not* equipped with a torque sensor, the underlying “torque control loop” is implemented as feedforward control being severely sensitive to parameter variations and uncertainties. Hence, the generator torque might drastically *differ* from the required (reference) torque and affect annual power production and gross earnings.

Many high-level control strategies assume that (i) the actual torque equals the demanded torque *instantaneously* (neglecting the dynamics of the electrical system, see [2], [3]) and *accurately* (neglecting possible deviations between demanded and actual torque) or assume that (ii) the machine torque is proportional to the difference between synchronous speed and machine speed (which only holds in steady state, see [4]).

†All authors (in alphabetic order) contributed equally to this paper.

Torque deviations violate this assumption and directly affect the power production of modern WTS. Hence, for benchmark tests, faults leading to torque deviations are considered to have a medium to high impact on the power production, see [5].

## II. Causes for non-ideal torque control

The possible causes for non-ideal torque control are illustrated in Fig. 1 and are due to

- *sensor faults* and *measurement errors* (↯) during the data acquisition of e.g.
  - the wind speed $v_{W,\text{meas}}\,[\text{m/s}]$,
  - the rotational speed $\omega_{M,\text{meas}}\,[\text{rad/s}]$,
  - the machine currents $\boldsymbol{i}_{M,\text{meas}}\,[\text{A}]^3$,
- *communication delays* and *signal processing errors* (↯) during data handling of e.g.
  - the torque reference $m_{M,\text{ref}}\,[\text{Nm}]$,
  - the voltage reference $\boldsymbol{u}_{M,\text{ref}}\,[\text{V}]^3$
- and, mainly, *parameter variations* and *parameter uncertainties* (↯) in certain components, such as
  - the converter,
  - the machine, and
  - the turbine (including friction).

The different causes can have different impact on torque control. Deviations between reference torque $m_{M,\text{ref}}\,[\text{Nm}]$ and actual machine torque $m_M\,[\text{Nm}]$ can manifest themselves as (i) a constant or varying offset or (ii) a constant or varying scaling of the desired (reference) torque or (iii) a combination of both. Since, in this paper, only the effects *during* energy production are considered (i.e. $m_M \neq 0$), the occurring torque deviations $m_M = \gamma\, m_{M,\text{ref}}$ can be modeled by a simple multiplication with the *deviation factor* $\gamma$ [1]. For simplicity, we only consider a constant factor, i.e.

**Assumption (A.1)** *The actual machine torque* $m_M\,[\text{Nm}]$ *(a non-linear function of flux linkage and machine currents) and reference torque* $m_{M,\text{ref}}\,[\text{Nm}]$ *differ by a constant* deviation factor $\gamma$*, i.e.*

$$\boxed{\forall\, \gamma \in (0, \gamma_{\max})\ \forall\, t \geq 0: \qquad m_M(t) = \gamma\, m_{M,\text{ref}}(t).} \tag{1}$$

Moreover, for the upcoming analysis, we impose the following assumptions

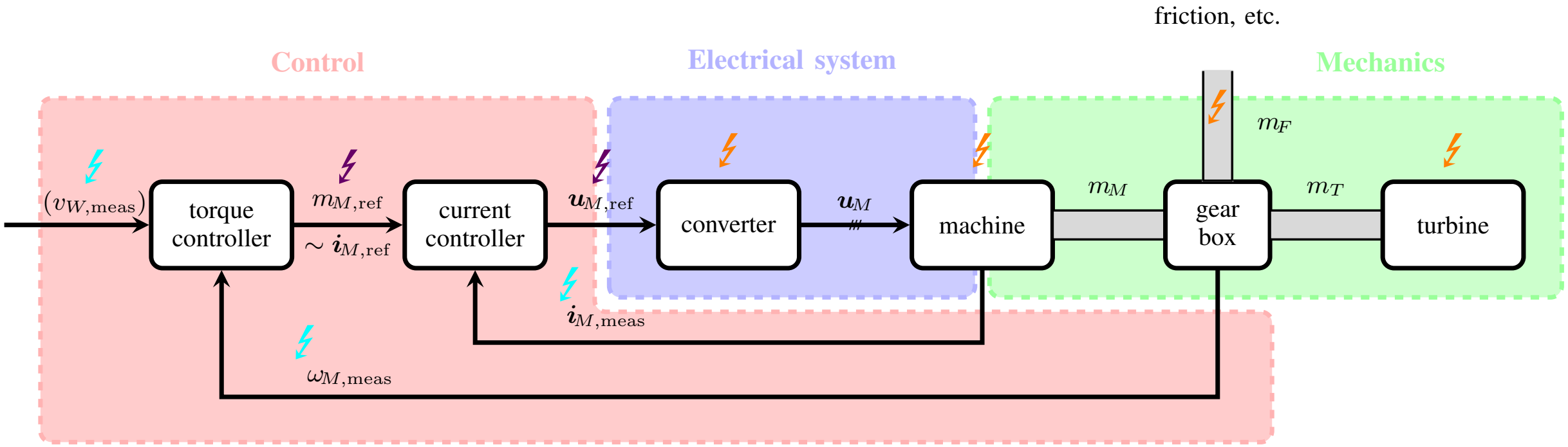

Figure 1: *Causes for non-ideal torque control.*

**Assumption (A.2)** *The pitch control system is ideal (e.g. no dynamics and deviations), i.e.*

$$\beta(t) = \beta_{\text{ref}}(t), \qquad \forall\, t \geq 0, \tag{2}$$

*with pitch angle $\beta\,[°]$ and its reference value $\beta_{\text{ref}}\,[°]$.*

**Assumption (A.3)** *All losses in the wind turbine systems are neglected; e.g. friction (in the mechanical drive train), iron and copper losses in the electrical machine and conducting and switching losses in the converters are* not *considered.*

## III. Operation, modeling and control of wind turbine systems with non-ideal torque control

In this section, the principles of wind turbine operation are explained, a simple turbine model is introduced and the control objectives and control methods for the different operation regimes are re-visited. Finally, the impact of non-ideal torque control is discussed for the four operation regimes.

### A. Regimes of operation

Depending on the actual wind speed $v_W$ [m/s], the wind turbine system will operate in one of four regimes of operation (see Fig. 2). For too less or too much wind (i.e. $v_W < v_{\text{cut-in}}$ in regime I and $v_W \geq v_{\text{cut-out}}$ in regime IV, resp.) the wind turbine is (usually[1]) at standstill or in idle speed: The turbine angular velocity is zero, i.e. $\omega_T = 0\,\frac{\text{rad}}{\text{s}}$ or the machine torque is zero, i.e. $m_M = 0\,\text{N m}$, hence $p_T = 0\,\text{W}$.

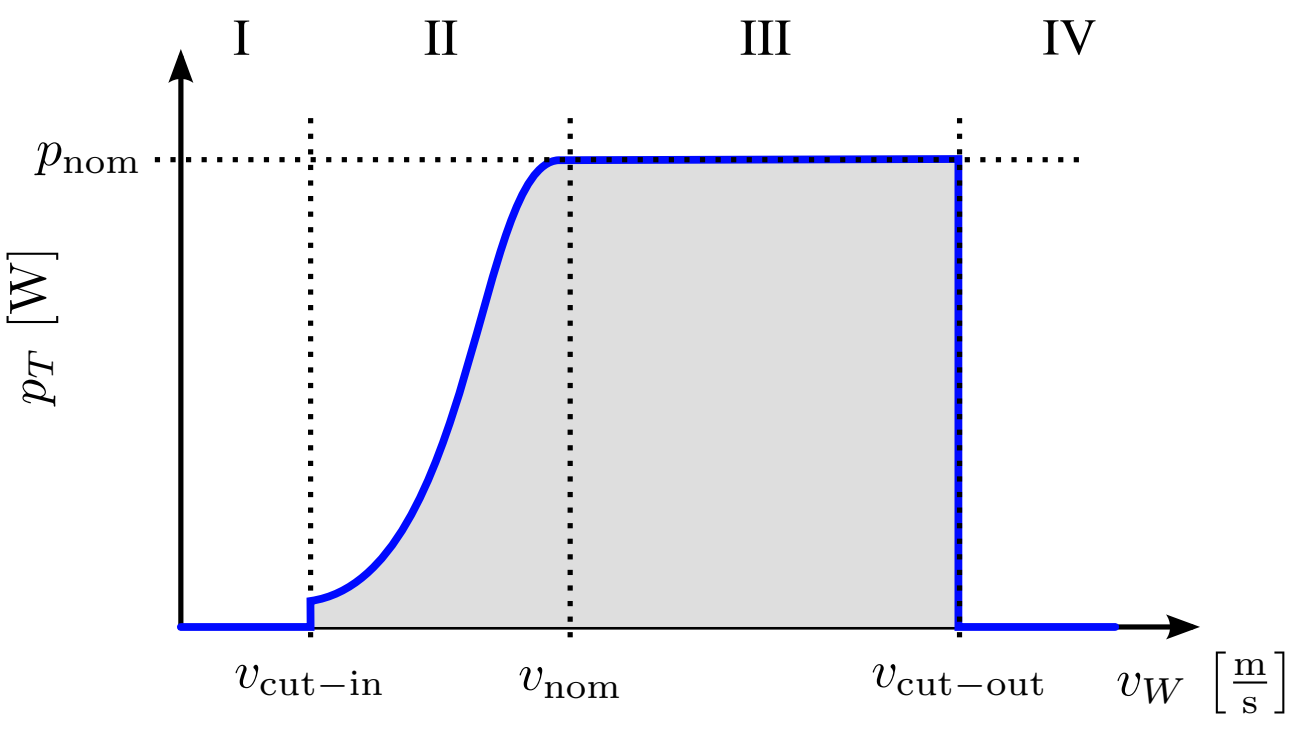

Figure 2: *Operation regimes of a WTS:*
- *Regime I: Standstill (too less wind), i.e. $p_T = 0$,*
- *Regime II: Variable power, i.e. $0 \leq p_T < p_{T,\text{nom}}$ (Goal: Maximum power point tracking),*
- *Regime III: Nominal power, i.e. $p_T = p_{T,\text{nom}}$,*
- *Regime IV: Standstill (too much wind), i.e. $p_T = 0$.*

In regime II, the wind speed is below the nominal wind speed $v_{\text{nom}}$ [m/s] but above the (minimum) cut-in wind speed $v_{\text{cut-in}}$ [m/s]. Due to the time-varying nature of the wind speed $v_W(\cdot)$, the turbine output power will vary between zero and nominal power $p_{T,\text{nom}}$ [W], i.e. $0 \leq p_T < p_{T,\text{nom}}$. The goal is to extract as much wind power as possible, i.e. *maximum power point tracking* (MPPT) which is achieved by an underlying speed controller (see Sec. III-C2).

In regime III, the wind speed is higher than the nominal wind speed but lower than the (maximum) cut-out wind speed $v_{\text{cut-out}}$ [m/s], i.e. $v_{\text{cut-out}} > v_W \geq v_{\text{nom}}$, and the torque $m_M$ of the electrical machine is kept constant at its nominal value (by constant feedforward torque control and ideal pitch control, see Assumption (A.2)). The nominal power output is generated, i.e. $p_T = p_{T,\text{nom}}$ (see Sec. III-C3).

### B. Simplified wind turbine model (see [2], [7] or [8, Ch. 8])

The wind power is given by

$$p_W(t) = \frac{1}{2}\rho\pi r_T^2 v_W^3(t)\,[\text{W}] \tag{3}$$

with air density $\rho\,[\text{kg/m}^3]$, turbine radius $r_T$ [m], and wind speed $v_W$. The extractable power can be approximated by introducing a power coefficient $c_P(\lambda, \beta)$ [1] which is a function of the tip speed ratio $\lambda := \frac{r_T \omega_T}{v_W}$ [1] (depending on turbine radius $r_T$ [m], turbine rotational speed $\omega_T$ [rad/s] and wind speed $v_W$) and the pitch angle $\beta$. The (mechanical) turbine power

$$p_T(\lambda, \beta, v_W) = c_P(\lambda, \beta)\, p_W(t)\,[\text{W}], \tag{4}$$

is limited by the Betz limit with $p_T(\lambda, \beta, v_W) \leq \frac{16}{27} p_W(t)$, see [9]. From turbine power (5), the turbine torque

$$m_T(\lambda, \beta, \omega_T) = \frac{p_T(\lambda, \beta, v_W)}{\omega_T} \overset{(3)}{\underset{(4)}{=}} \frac{1}{2}\rho\pi r_T^2 v_W^3 \frac{c_P(\lambda, \beta)}{\omega_T} = \frac{1}{2}\rho\pi r_T^5 \frac{c_P(\lambda, \beta)}{\lambda^3}\omega_T^2 \;[\text{Nm}] \tag{5}$$

can be derived as function of the rotational speed $\omega_T$ [rad/s]. Note that (5) is only a reasonable model for positive wind speeds and angular speeds [10] (for $\omega_T = 0$, (5) gives a zero torque which is physically not reasonable).

[1] Some companies use more sophisticated control methods for high winds, e.g. see patent [6] of REpower Systems for a reduced power production above $v_{\text{cut-out}}$ instead of a shut down.

Considering Assumption (A.3) and

**Assumption (A.4)** *The mechanical coupling in the shaft is stiff and the (ideal) gear box has gear ratio* $g_r > 1$ *[1],*

the simplified dynamics of the mechanical drive train of the wind turbine systems can be modeled by [2]

$$\frac{\mathrm{d}}{\mathrm{d}t}\omega_M(t) = \frac{1}{\Theta}\left(\frac{m_T(\lambda,\beta,\omega_T)}{g_r} + m_M(t)\right) \qquad (6)$$

with angular machine velocity $\omega_M = g_r\omega_T$ [rad/s], (total) inertia $\Theta$ [kgm$^2$], turbine torque $m_T$ and machine torque $m_M$.

### C. Wind turbine control methods for the different regimes

*1) Regime I and IV:* In regime I and IV, torque control is *not* active and no power is produced, i.e. $p_T = 0$ (see Sec. III-A). Hence, an analysis of the impact of non-ideal torque control on the power production is not meaningful.

*2) Regime II:* We consider the MPPT strategy proposed in [7] or [8, Ch. 8] which assures throughout regime II that

(i) the pitch angle of the turbine blades is kept at its optimum $\beta_{\text{opt}}$ [°], i.e. $\beta_{\text{ref}} = \beta = \beta_{\text{opt}}$, and
(ii) the tip speed ratio is kept at its optimum $\lambda^\star_{\text{opt}}$ [1], i.e. $\lambda = \lambda^\star_{\text{opt}}$ such that the turbine operates with its maximal power coefficient $c^\star_{P,\text{opt}} := c_P(\lambda^\star_{\text{opt}}, \beta_{\text{opt}})$ [1] (see red curve in Fig. 4).

Figure 3 illustrates the control strategy in regime II. To guarantee that the wind turbine system operates at $\lambda^\star_{\text{opt}}$, the following nonlinear speed controller

$$m_{M,\text{ref}}(t) = -\underbrace{\frac{1}{2}\frac{\rho\pi r_T^5}{g_r^3}\frac{c_P(\lambda^\star_{\text{opt}},\beta_{\text{opt}})}{\left(\lambda^\star_{\text{opt}}\right)^3}}_{=:k^\star_P}\omega_M(t)^2 \qquad (7)$$

has been proposed (for details on stability of the closed-loop system dynamics, see [2], [7] or [8, Ch. 8]). Note that the closed-loop system (6), (7) reaches and remains in its equilibrium if and only if

$$\forall t \geq 0: \qquad m_M(t) = -\frac{m_T(\lambda(t),\beta(t),\omega_T(t))}{g_r}. \qquad (8)$$

Analyzing (6), (7) in steady state (i.e. $\frac{\mathrm{d}}{\mathrm{d}t}\omega_M = 0$) gives

$$\begin{aligned} 0 &= \frac{m_T(\lambda,\beta,\omega_T)}{g_r} + m_M \\ &\overset{(1)}{=} \frac{m_T(\lambda,\beta,\omega_T)}{g_r} + \gamma\, m_{M,\text{ref}} \qquad (9) \\ &\overset{(5)}{\underset{(7)}{=}} \frac{1}{2}\frac{\rho\pi r_T^5}{g_r^3}\left(\frac{c_P(\lambda,\beta)}{\lambda^3} - \gamma\frac{c_P(\lambda^\star_{\text{opt}},\beta_{\text{opt}})}{\left(\lambda^\star_{\text{opt}}\right)^3}\right)\omega_M^2 \qquad (10) \end{aligned}$$

which shows that for $\lambda = \lambda^\star_{\text{opt}}$, $\beta = \beta_{\text{opt}}$, and $\gamma = 1$ (ideal torque control) the wind turbine system operates at its maximum power point (MPP, see Fig. 4), i.e. $(\lambda, c_P(\lambda,\beta)) \overset{!}{=} (\lambda^\star_{\text{opt}}, c^\star_{P,\text{opt}}) = (\lambda^\star_{\gamma=1.0}, c^\star_{P,\gamma=1.0})$.

*Discussion of the impacts of non-ideal torque control:* What happens for non-ideal torque control, i.e. $\gamma \neq 1$ in (10)? Clearly, assuming that the closed-loop system (6), (7) remains stable, it will *not* converge to the MPP. The impact of $\gamma \neq 1$ on the power coefficient is illustrated in Fig. 4: For $\gamma > 1$, the equilibrium is shifted to the left and, for $\gamma < 1$, it is shifted to the right of the MPP. In general, the attained (steady state) tip speed ratio will differ from its optimal value, i.e. $\lambda^\star \neq \lambda^\star_{\text{opt}}$, and, hence, the wind turbine operates with a reduced power coefficient, i.e. $c^\star_{P,\gamma\neq 1} < c^\star_{P,\text{opt}} = c^\star_{P,\gamma=1.0}$ (see Fig. 4). A more detailed analysis of the dynamical behavior of the closed-loop system (6), (7) can be found in [2].

Concluding, for regime II and non-ideal torque control, the turbine power $p_T$ is decreased and (in steady state) given by

$$\boxed{p_T = \frac{1}{2}\rho\pi r_T^2\, c^\star_{P,\gamma}\, v_W^3} < \frac{1}{2}\rho\pi r_T^2\, c^\star_{P,\text{opt}}\, v_W^3. \qquad (11)$$

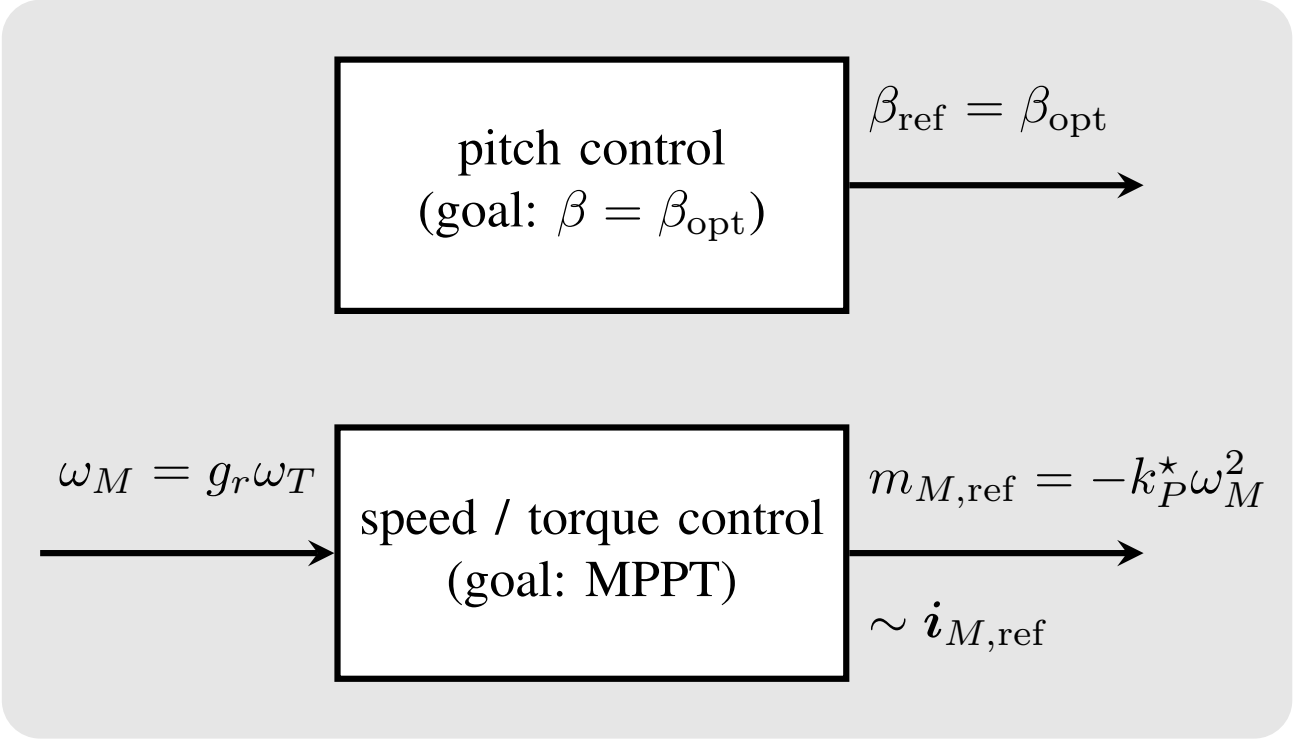

Figure 3: *Control strategy for regime II.*

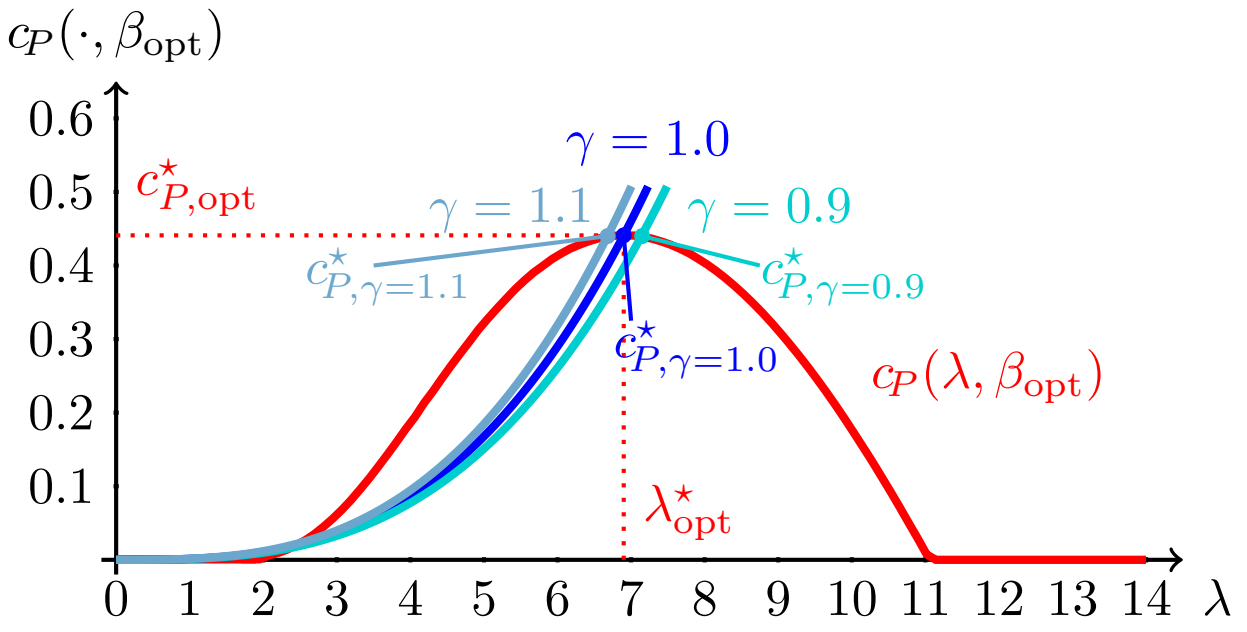

Figure 4: *Power coefficient of WTS and deviations due to* $\gamma$*:*

- $0.9\frac{c_P(\lambda^\star_{\text{opt}},\beta_{\text{opt}})}{(\lambda^\star_{\text{opt}})^3}\lambda^3$
- $1.0\frac{c_P(\lambda^\star_{\text{opt}},\beta_{\text{opt}})}{(\lambda^\star_{\text{opt}})^3}\lambda^3$
- $1.1\frac{c_P(\lambda^\star_{\text{opt}},\beta_{\text{opt}})}{(\lambda^\star_{\text{opt}})^3}\lambda^3$

**Remark III.1.** *The presented state-of-the-art control strategy in regime II can also be extended by inertia compensation to achieve faster closed-loop system dynamics (see [8, Ch. 8]). However, this will not affect our analysis of the impact of non-ideal torque control yielding operation* away *from the MPP.*

*3) Regime III:* In regime III, the control objective is to limit the extracted power to its rated value. The rotational machine speed $\omega_M = g_r\omega_T$ is limited to its nominal value $\omega_{M,\text{nom}} = g_r\omega_{T,\text{nom}}$ [rad/s] by adequate pitching of the turbine blades.

Hence, the speed control is now shifted to the pitch control system, whereas the machine torque is kept constant at its rated value $m_M = m_{M,\mathrm{nom}} = \frac{m_{T,\mathrm{nom}}}{g_r}$. So, it is assured that the nominal power

$$p_{T,\mathrm{nom}} = \omega_{T,\mathrm{nom}} m_{T,\mathrm{nom}} = \omega_{M,\mathrm{nom}} m_{M,\mathrm{nom}} \qquad (12)$$

is extracted (see [8, Ch. 8]). Figure 5 illustrates the control strategy for regime III.

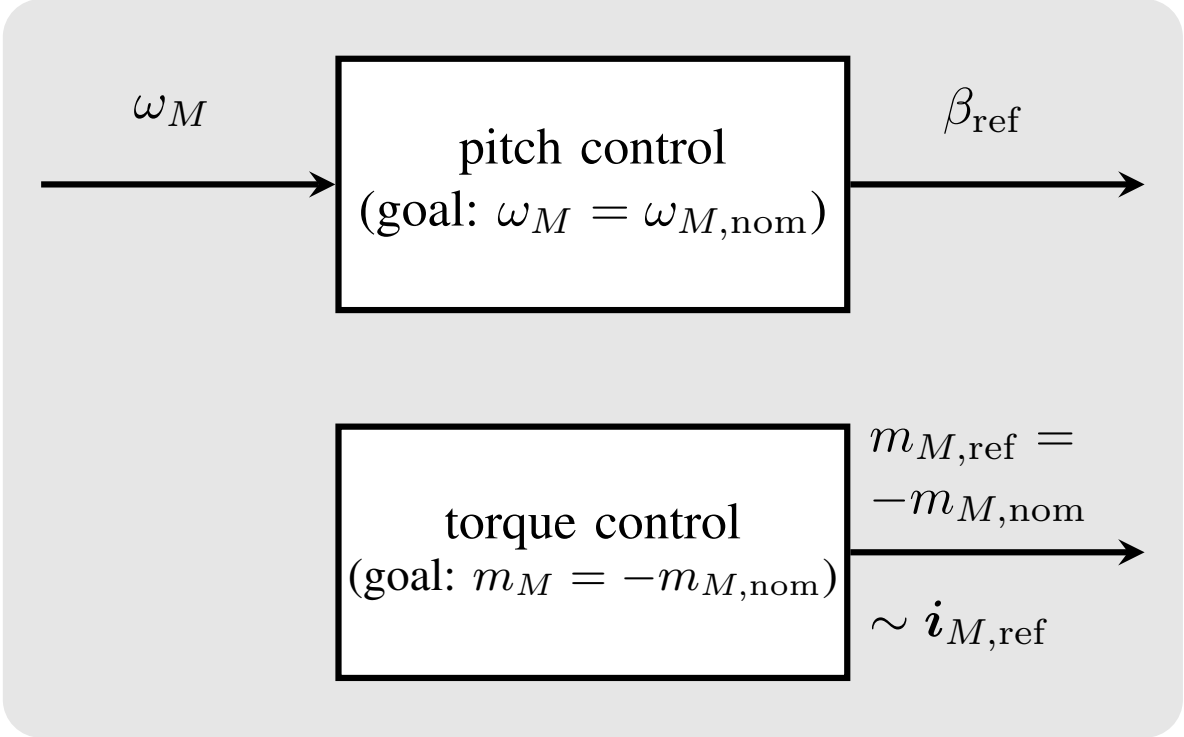

Figure 5: *Control strategy for regime III.*

*Discussion of the impact of non-ideal torque control:* What influence on the power production is to be expected for $\gamma \neq 1$? First note that, in steady state and in view of Assumption (A.2) (ideal pitch control system), the rotational speed is kept constant at its nominal value, i.e.

$$\omega_T = \omega_{T,\mathrm{nom}} \quad \Longleftrightarrow \quad \omega_M = \omega_{M,\mathrm{nom}}. \qquad (13)$$

Moreover, control strategy in regime III assures

$$m_{M,\mathrm{ref}} = -m_{M,\mathrm{nom}}, \qquad (14)$$

which gives in steady state the following deviation from the nominal turbine power

$$p_T \overset{(13)}{=} \omega_{T,\mathrm{nom}} m_T \overset{(9)}{=} -\gamma\, \omega_{M,\mathrm{nom}} m_{M,\mathrm{ref}} \overset{(14)}{=} \gamma\, \omega_{M,\mathrm{nom}} m_{M,\mathrm{nom}}$$
$$\Longrightarrow \boxed{p_T \overset{(12)}{=} \gamma\, p_{T,\mathrm{nom}}.} \qquad (15)$$

**Remark III.2.** *There exists also a slightly different approach to control the turbine power (see [8, Ch. 8]): The pitch controller still keeps the rotational speed at its nominal value. Due to the slow dynamics of the pitch control system, the actual rotational speed might differ from its nominal value. To keep the turbine power output constant at its nominal value, the reference machine torque* $m_{M,\mathrm{ref}}(t) \overset{(8)}{=} -\frac{p_{T,\mathrm{nom}}}{g_r \omega_{T,\mathrm{meas}}(t)}$ *is adjusted online based on the actual measurement of the turbine rotational speed* $\omega_{T,\mathrm{meas}}\,[\mathrm{rad/s}]$. *Nevertheless, this control method is also affected by non-ideal torque control.*

### D. Other control strategies

A few control strategies in the literature are able to overcome the problems of non-ideal torque control (although *not* explicitly addressing the problem). In [11] a torque observer for a permanent magnet synchronous generator is designed using current measurements. Another torque observer is proposed in [12] for condition monitoring. If exact knowledge of the power curve/coefficient of the turbine is available, deviations in turbine speed due to non-ideal torque control can be compensated for in region II (see [13]). Another option for region II is to use the measured or estimated value of the wind speed to compute and adjust the actual machine/turbine reference speed online (see [10, 14, 15]). In [16] the use of a turbine-mounted LIDAR system can improve the power output in regime II. In [17] a nonlinear torque controller for regime III is proposed which utilizes the actual value of the electrical output power and the rotational speed of the turbine. Most of the approaches are still sensitive to parameter uncertainties and variations. Next, we will discuss the impact of non-ideal torque control on the power output and the corresponding annual losses in the gross earnings for a turbine operator.

Table I: *System, controller and financal data.*

| description | symbols & values (with unit) |
|---|---|
| *System data* | |
| rotor radius | $r_T = 40\,\mathrm{m}$ |
| air density | $\rho = 1.293\,\frac{\mathrm{kg}}{\mathrm{m}^3}$ |
| gear ratio | $g_r = 1$ |
| power coefficient | $c_P(\lambda, \beta) = c_{P,2}(\lambda, \beta)$ as in [2] |
| *Controller data* | |
| wind speed tresholds | $v_{\mathrm{cut-in}} = 4\,\frac{\mathrm{m}}{\mathrm{s}}$ |
| for regimes I, …, IV | $v_{\mathrm{nom}} = 11\,\frac{\mathrm{m}}{\mathrm{s}}$ |
| | $v_{\mathrm{cut-out}} = 25\,\frac{\mathrm{m}}{\mathrm{s}}$ |
| power coefficient | $c^\star_{P,\gamma=1.0} = 0.441$ |
| (computed numerically | $c^\star_{P,\gamma=0.9} = 0.439$ |
| based on the dynamical | $c^\star_{P,\gamma=0.8} = 0.433$ |
| WTS model in [2]) | $c^\star_{P,\gamma=0.7} = 0.423$ |
| *Financial data* | |
| feed-in tariff | $0.08\,\frac{€}{\mathrm{kW\,h}}$ |

## IV. Simulation results

For the upcoming simulations, the variable-speed pitch-controlled $2\,\mathrm{MW}$ wind turbine system presented in [2] is considered. To compute the annual energy production of the wind turbine system, real one-year wind speed measurements with a time resolution of $10\,\mathrm{min}$ from FINO1[2] are used. Simulation, controller and financial data are collected in Tab. I.

### A. Computation of annual power and energy

For the computation of the turbine power in regimes II and III, the steady state formulas (11) and (15) for the turbine power are evaluated. For regime I and IV, we set $p_T = 0$.

Hence, the generated energy in each operation regime $\alpha \in \{I, II, III, IV\}$ can be computed for every $10\,\mathrm{min}$ by multiplying the respective turbine power $p_T^\alpha$ with weighting factor $k^{\alpha} \in \{0,1\}$ and time interval 600 s. A year can be divided into $n \in \{1,\dots,52560\}$ time intervals of 10 min = 600 s length each. Depending on the actual wind speed $v_W[n]$, turbine power $p_T[n]$ and energy $E_T[n]$ are calculated for all $n \in \{1,\dots,52560\}$ as follows (iteration over all $n$):

- Regime I (**if** $0 \le v_W[n] < v_{\text{cut-in}}$ **then** $k^I[n] = 1$, **else** $k^I[n] = 0$): $p_T^I = 0$
$$\implies E_T^I[n] := p_T^I \cdot k^I[n] \cdot 600\,\text{s} = 0; \tag{16}$$
- Regime II (**if** $v_{\text{cut-in}} \le v_W[n] < v_{\text{nom}}$ **then** $k^{II}[n] = 1$, **else** $k^{II}[n] = 0$): $p_T^{II}[n] = c^{\star}_{P,\gamma} \frac{1}{2}\rho\pi r_T^2 v_W[n]^3$
$$\implies E_T^{II}[n] := p_T^{II}[n] \cdot k^{II}[n] \cdot 600\,\text{s}; \tag{17}$$
- Regime III (**if** $v_{\text{nom}} \le v_W[n] < v_{\text{cut-out}}$ **then** $k^{III}[n] = 1$, **else** $k^{III}[n] = 0$): $p_T^{III} = \gamma\, p_{T,\text{nom}}$
$$\implies E_T^{III}[n] := p_T^{III} \cdot k^{III}[n] \cdot 600\,\text{s}; \tag{18}$$
- Regime IV (**if** $v_W[n] \ge v_{\text{cut-out}}$ **then** $k^{IV}[n] = 1$, **else** $k^{IV}[n] = 0$): $p_T^{IV} = 0$
$$\implies E_T^{IV}[n] := p_T^{IV} \cdot k^{IV}[n] \cdot 600\,\text{s} = 0. \tag{19}$$

[2]The wind data was measured at the research platform FINO1 (with geographical coordinates: $54^\circ\ 00'\ 53,5''\,N$, $06^\circ\ 35'\ 15,5''\,E$) between 24[th] November 2012 and 24[th] November 2013. Mean values were saved with a $10\,\mathrm{min}$ resolution.

The total (annual) energy production (in [Wh])

$$E_T := \sum_{n=1}^{52560} E_T^I[n] + E_T^{II}[n] + E_T^{III}[n] + E_T^{IV}[n]$$

is given by the sum of the energies of all regimes.

### B. Discussion of the simulation results

The simulation results for ideal torque control (i.e. $\gamma = 1$) and non-ideal torque control ($\gamma \neq 1$) are shown in Fig. 6. All simulations are fed by the measured wind speed profile from FINO1. For non-ideal torque control, three different scenarios are investigated with three different deviation factors $\gamma \in \{0.9,\ 0.8,\ 0.7\}$.

The first subplot in Fig. 6a shows the measured wind speed over one year. The thresholds $v_{\text{cut-in}}$, $v_{\text{nom}}$ and $v_{\text{cut-out}}$ for the different operation regimes are marked by -·-·-, ——, and - - -, respectively. The resulting turbine power $p_T$ is shown in the second subplot. The third subplot depicts the evolution of the total energy $E_T$ over time. For decreasing values of the deviation factor $\gamma$, the impact of non-ideal torque control is more significant; turbine power and energy reduce more and more severely.

Figures 6b and 6c show the total energy[3] and the corresponding revenue for a feed-in tariff of $0.08\,\frac{€}{\text{kW h}}$, respectively. For $\gamma = 0.9$, the energy production is reduced by 0.601 GW h (or 6.6 %). For $\gamma = 0.7$, it is already reduced by 1.891 GW h (or 20.7 %). The energy reductions correspond to annual earning losses of (i) 48 080 € for $\gamma = 0.9$, (ii) 98 480 € for $\gamma = 0.8$ and (iii) 151 280 € for $\gamma = 0.7$ (see also Tab. II).

[3] In view of Assumption (A.3), the turbine energy $E_T$ equals the produced electrical energy $E_{\text{elec}}$ [Wh], i.e. $E_T = E_{\text{elec}}$.

Table II: *Comparison of earnings and losses.*

| | Energy | Earnings | % |
|---|---|---|---|
| earnings ($\gamma = 1.0$) | 9.122 GW h | 729 760 € | 100 % |
| losses ($\gamma = 0.9$) | 0.601 GW h | 48 080 € | 6.6 % |
| losses ($\gamma = 0.8$) | 1.231 GW h | 98 480 € | 13.5 % |
| losses ($\gamma = 0.7$) | 1.891 GW h | 151 280 € | 20.7 % |

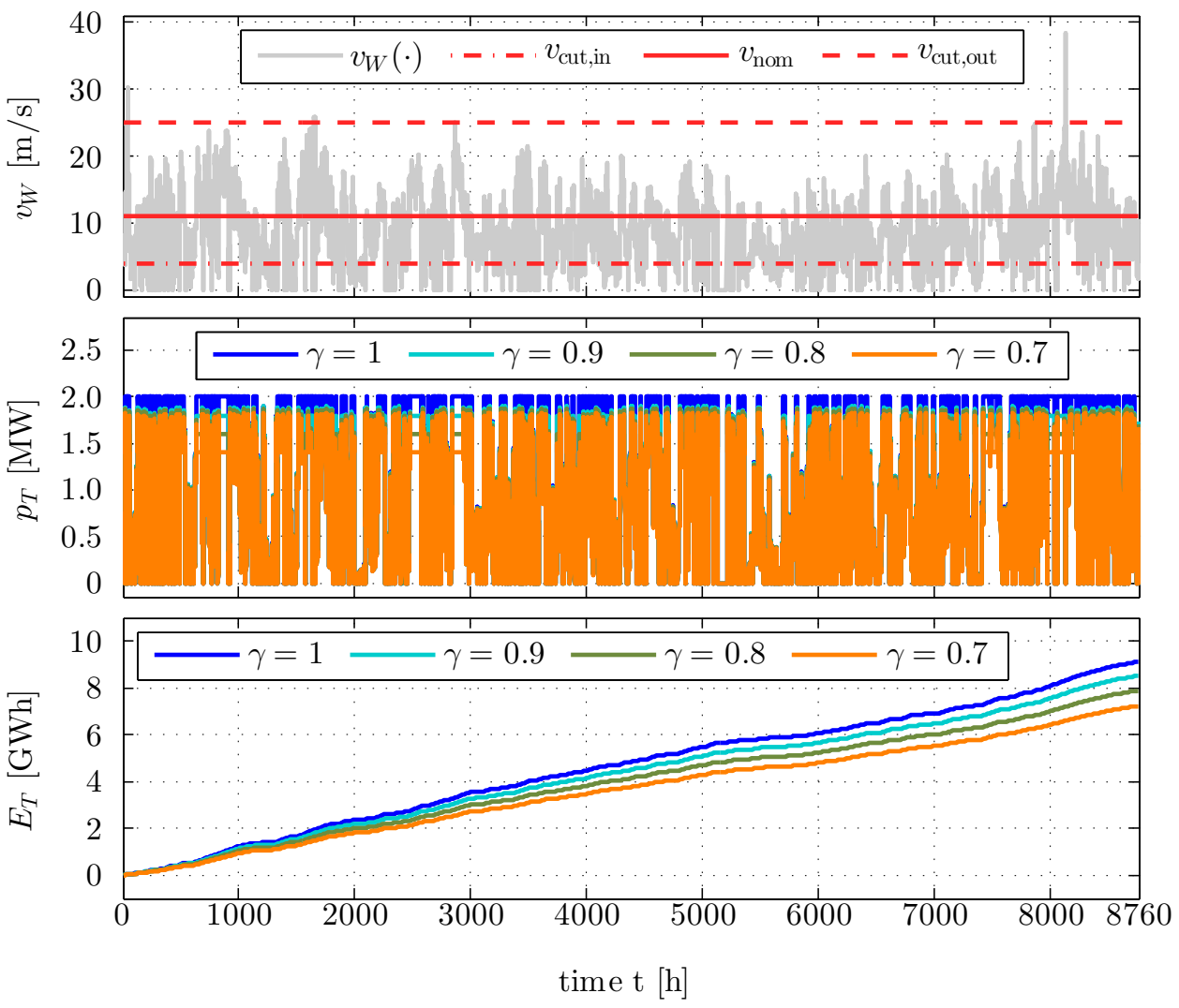

(a) *Annual wind profile, turbine power and energy production.*

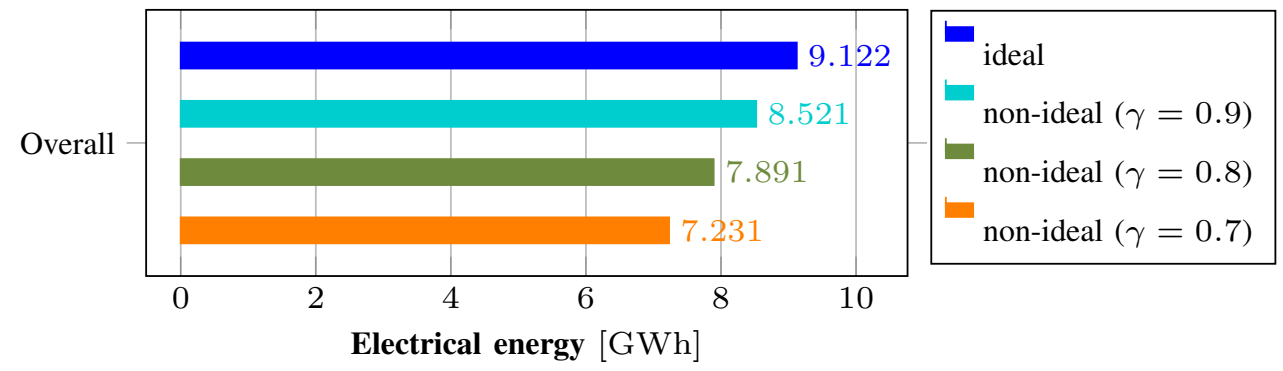

(b) *Comparison of electric energy production and losses.*

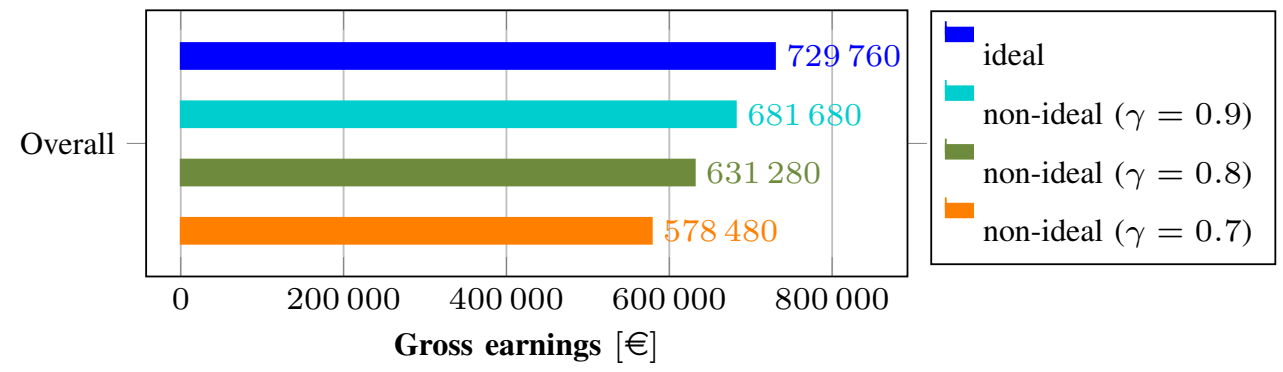

(c) *Comparison of gross earnings and losses.*

Figure 6: *Comparative simulation results for ideal ($\gamma = 1$) and non-ideal ($\gamma \in \{0.9,\ 0.8\,0.7\}$) torque control.*

Figure 7 illustrates the individual impacts of non-ideal torque control in regime II and III on the energy losses for the deviation factor $\gamma = 0.7$. The major share (i.e. 93 % of the overall annual losses) of the total losses are due to the impacts of non-ideal torque control in regime III. Since the graph of the power coefficient is quite flat near the MPP (see Fig. 4), the impacts of non-ideal torque control in regime II are less significant. The losses in region II only account for 7 % of the total losses.

**Remark IV.1.** *For the simulations only values smaller than one were considered, i.e. $\gamma < \gamma_{\max} \le 1$ were considered. Clearly, in regime III, a deviation factor of $\gamma > 1$ would*

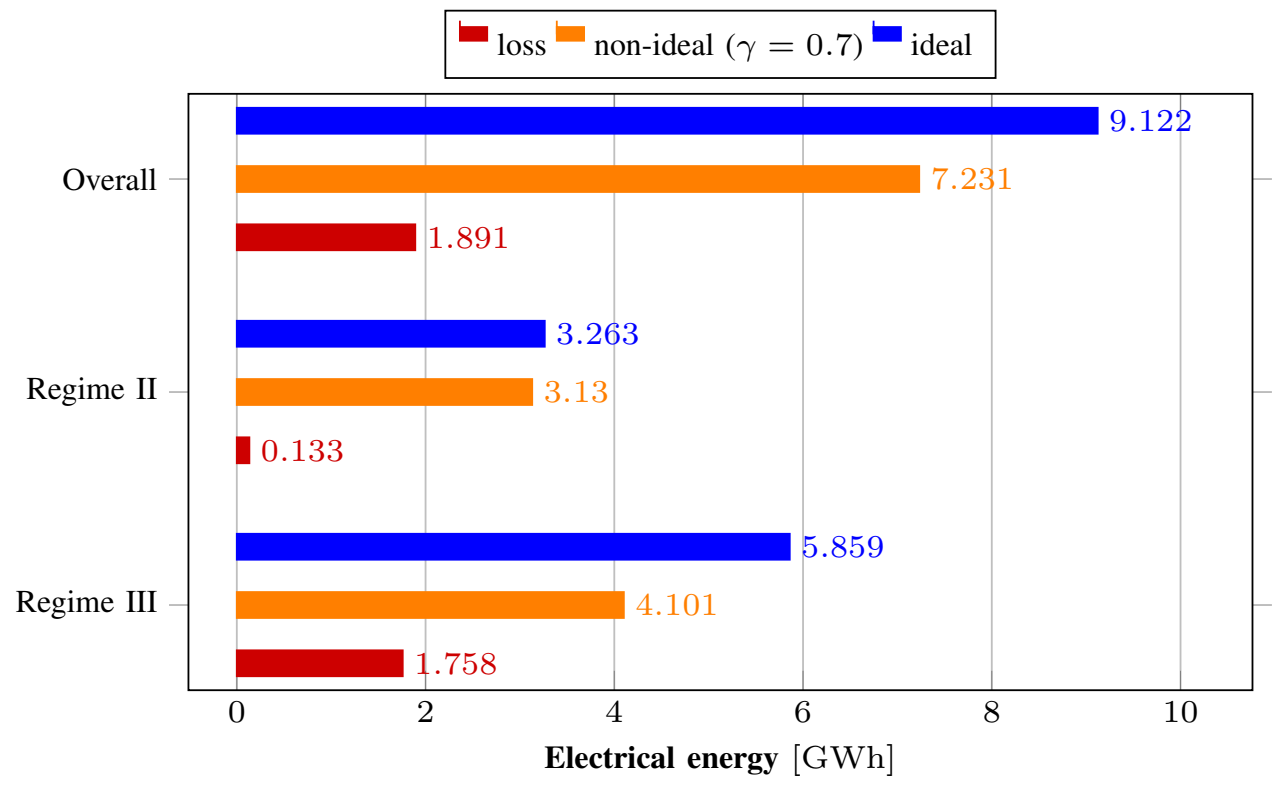

Figure 7: *Simulation results comparing ideal and non-ideal torque control ($\gamma = 0.7$) split up into different regimes.*

*result in a permanently higher power production than the rated power production. However, for this, the generator must apply a higher torque than the rated torque to the turbine which will cause damage to the wind turbine system (in particular, to the generator and/or the converter).*

## V. Conclusion

It has been shown that non-ideal torque control *is* a problem and may (drastically) reduce energy production of wind turbine systems (if no counter measures are taken). For state-of-the-art wind turbine control systems, the impact of non-ideal torque control on the power production during partial load (regime II) and full load (regime III) has been discussed. Exemplary simulations were performed for a variable-speed pitch-controlled 2 MW wind turbine system fed by a real wind speed profile (measured wind data over one year). The simulation results illustrate the drastic earning losses depending on the deviations between actual and reference machine torque. Although, a few papers propose more sophisticated control methods that would not be affected by this problem, there is still not a completely satisfying solution at hand. Necessary steps to avoid non-ideal torque control are (i) condition monitoring, (ii) fault detection and identification, (iii) online parameter estimation and error/fault compensation and (iv) advanced control techniques (e.g. non-linear and/or adaptive methods).

## Acknowledgement

The authors are deeply grateful to the FINO-Project (BMU, PTJ, BSH, DEWI GmbH) for providing the wind data used for the simulation.